\documentclass[sigconf]{acmart}
\acmConference[Preprint]{Preprint}{August 2025}{arXiv}

\renewcommand\footnotetextcopyrightpermission[1]{} 

\AtBeginDocument{%
  }

\usepackage{tikz}
\usepackage{pgfplots}
\usepackage{pgfplotstable}

\usepackage{dirtytalk}
\usepackage{booktabs}

\usepackage{cleveref}

\usepackage{xparse}

\usepackage{my-acronyms}

\usetikzlibrary{plotmarks}
\usetikzlibrary{calc}
\usetikzlibrary{patterns}

\usepgfplotslibrary{groupplots}
\pgfplotsset{compat=1.18}
\usepackage{pgfplots}
\usetikzlibrary{positioning, arrows.meta}
\usetikzlibrary{decorations.pathmorphing, positioning}

\begin{document}

\title{Back to Bits: Extending Shannon's communication performance framework to computing}

\author{Max Hawkins}
\orcid{1234-5678-9012}
\affiliation{%
  \institution{Georgia Institute of Technology}
  \city{Atlanta}
  \state{Georgia}
  \country{USA}
}
\email{mhawkins60@gatech.edu}

\author{Richard Vuduc}
\orcid{0000-0003-2178-138X}
\affiliation{%
  \institution{Georgia Institute of Technology}
  \city{Atlanta}
  \state{Georgia}
  \country{USA}
}
\email{richie@gatech.edu}

\renewcommand{\shortauthors}{Hawkins et al.}

\begin{abstract}
    We propose a novel measurement unit for computing performance, grounded in information theory.
    Modern computing systems are increasingly diverse, supporting low-precision formats, hardware specialization, and emerging paradigms such as analog, quantum, and reversible logic.
    Traditional metrics like floating-point operation (flop) counts no longer accurately capture this complexity.
    Instead, we frame computing as the transformation of information through a channel and define performance in terms of the mutual information between a system's inputs and outputs.
    This approach measures not just the quantity of data processed, but the amount of meaningful information encoded, manipulated, and retained through computation.
    Our framework provides a principled, implementation-agnostic foundation for evaluating performance.
\end{abstract}





\maketitle

\section{Introduction}\label{sec:intro}

Consider two different computers that report the same floating-point operations per second (flop/s).
To what degree does this signify equal performance?
Given the current variety of computing data types, operations, and hardware, the answer is unclear.

For instance, suppose the first computer reports its performance for 64-bit IEEE-754 floating point operations (FP64) while the other uses 8-bit Open Compute Project data types~\cite{rouhani_microscaling_2023}. 
Does it matter that an operation on 64-bit data types requires many more transistors and thus chip area and energy consumption compared to an equivalent operation with 8-bit data types?
Even comparing across similar bit widths, does it matter if the formats differ in their encoding schemes, value ranges, precision, or representable values?
Which operations are considered valuable?
Only multiplies and adds?
What if the operations are lossy/noisy?
How should these differences affect how we measure and compare the performance of these two platforms?

The essential problem, in our view, is the choice of unit of measure: the \emph{flop}~\cite{kuck_computer_1974}.
Counting flops as part of reporting performance is like using `floating-point elements' as a basis for communication throughput: both conflate format with function.
For the past 40 years, we mostly assumed that a flop referred to a specific number representation and choice of elementary operations (adds and multiplies in IEEE-754 FP64).
A flop count was a reasonable proxy for the intrinsic computational work of most core algorithms in scientific computing.
But now, this assumption is upended by the ongoing proliferation of new formats and hardware paradigms.
The classical notion of a flop no longer supports fair performance comparisons---qualifying flops with types or sparsity is at best cumbersome, and at worst, compounding the problem of performance comparison.

In fact, the U.S. government's computing export controls have codified this inadequacy for more than twenty years, applying changing corrective factors to flops: 
from Composite Theoretical Performance (CTP) in 1991, which applied a non-linear scaling based on data type bit width ~\cite{ames_composite_1991},
to the 2006 Weighted TeraFLOP (WT) metric emphasizing 64-bit arithmetic~\cite{us_department_of_commerce_practitioners_2006},
to the current metric, Total Processing Performance (TPP), which explicitly ignores sparsity and re-established a bit width scaling, at first in 2022 based on the maximum of either the inputs or outputs ~\cite{us_department_of_commerce_implementation_2022} before quick revision to considering only the input data type bit widths~\cite{us_department_of_commerce_implementation_2023, us_department_of_commerce_framework_2025}.
%
These changing metrics reflect the difficulty in evaluating computing performance without a principled foundation.

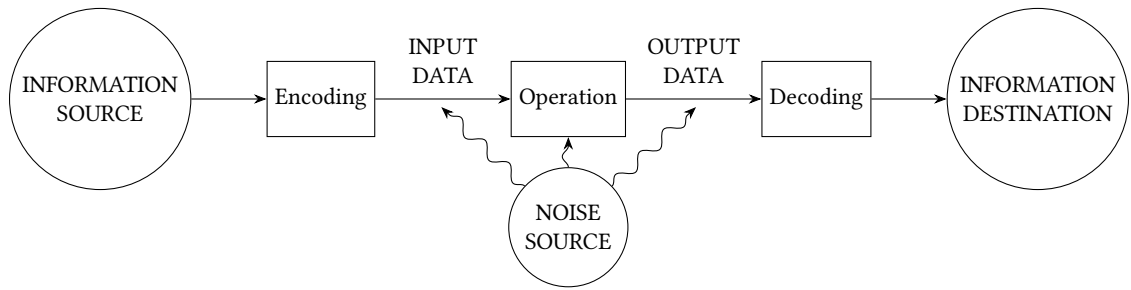
\begin{figure*}
    \centering
    \begin{tikzpicture}[
  node distance=1cm and 1.5cm,
  every node/.style={draw, minimum width=1.3cm, minimum height=1cm, align=center},
  >=Stealth,
  snake/.style={->, decorate, decoration={snake, amplitude=0.8mm, segment length=4.5mm}}
]

\node (source)[circle] {INFORMATION\\SOURCE};
\node[right=of source, right=1cm of source] (transmitter) {Encoding};
\node[right=of transmitter, right=1.8cm of transmitter] (box1) {Operation}; 
\node[below=of box1, circle, below=0.4cm of box1] (noise) {NOISE\\SOURCE};
\node[right=of box1, right=1.8cm of box1] (receiver) {Decoding};
\node[right=of receiver, circle, right=1cm of receiver] (destination) {INFORMATION\\DESTINATION};

\draw[->] (source) -- (transmitter);
\draw[->] (transmitter) -- node[above, draw=none] {INPUT\\DATA} (box1);
\draw[->] (box1) -- node[above, draw=none] {OUTPUT\\DATA} (receiver);
\draw[->] (receiver) -- (destination);

\draw[snake] (noise) -- (box1);
\draw[->, decorate, decoration={snake, amplitude=0.8mm, segment length=4.4mm}] (noise) -- ([xshift=16mm,yshift=4mm]transmitter.south);
\draw[->, decorate, decoration={snake, amplitude=0.8mm, segment length=4.4mm}] (noise) -- ([xshift=-16mm,yshift=4mm]receiver.south);

\end{tikzpicture}
    \caption{
    A generalized computing channel based on Shannon's communication model.
    Here, encoding/decoding is interpreted as data-type mappings into and out of computable (binary) form, while the channel represents the operation itself.
    }
    \label{fig:shannon_channel_capacity}
\end{figure*}

To address this, we propose a framework grounded in Shannon's information theory~\cite{shannon_mathematical_1948}. 
Just as a communication channel transforms inputs to outputs, transmitting information limited by its channel capacity, we model computation as a transformation channel that encodes, manipulates, and retains information (\cref{fig:shannon_channel_capacity}).
The mutual information between inputs and outputs defines the channel capacity of a computation.
This quantity, measured in bits of information per channel use, facilitates comparison across bit widths, data types, operations, and potentially hardware paradigms, making it a semantically-neutral and implementation-agnostic performance metric.
Beyond making performance reporting more uniform,
this framework can inform data type design and observed system utilization, while offering a basis to connect communication and computation analysis.

Our application of channel capacity thus offers a principled and forward-compatible foundation for measuring computational work---designed to capture the computing diversity of today and accommodate the architectures of tomorrow.

\section{Related Work}\label{sec:related_works}

The proposal most similar to ours is Ryabko's and Fionov's work on an instruction-based information theory performance metric~\cite{ryabko_estimating_2012}.
They also adopt Shannon's channel capacity metric~\cite{shannon_mathematical_1948}, but there is a critical distinction: they focus on instruction-set complexity whereas we focus on input and output \emph{data}.
Their work's \emph{instruction} centrism overlooks the role of information encoding, manipulation, and retention and thus does not directly solve the challenges presented by diverse data types and hardware.
In contrast, our approach returns to Shannon's original formulation of channel capacity, grounding computing performance in the mutual information between a operation's \emph{inputs and outputs}~\cite{shannon_mathematical_1948}.

Beyond instruction-centric views, Landauer's principle and thermodynamic analyses link computation to information loss~\cite{landauer_irreversibility_1961,bennett_thermodynamics_1982,wolpert_stochastic_2019}, while other work relates computational work to output entropy~\cite{cheng_entropy_1990}.
In machine learning, mutual information underpins the information bottleneck~\cite{tishby_deep_2015} and multilayer network optimization~\cite{linsker_self-organization_1988}.
Mackay explores similar ideas without quantifying performance~\cite{mackay_information_2003}, while Tononi and Sporns's `effective information' uses mutual information to quantify the strength of causal interactions between two systems~\cite{tononi_measuring_2003}.
Kolmogorov complexity offers another potential metric, but is much harder to use practically~\cite{kolmogorov_tables_1963,grunwald_shannon_2004}.
Quantum extensions (e.g., Holevo's work ~\cite{holevo_bounds_1973,holevo_capacity_1998,holevo_quantum_2012}) and photonic channels~\cite{lloyd_capacity_1997, amaolo_maximum_2024} further demonstrate information theory's applicability to emerging paradigms.

\section{Our Proposal: Computational Work as Mutual Information}\label{sec:proposal}


We suggest framing computation as a generalization of a communication channel.
Rather than simply copying data, a computer encodes information into data types, performs complex operations (beyond the identity), then decodes the outputs.
In his 1948 paper, Shannon abstracted communication away from message semantics~\cite{shannon_mathematical_1948}.
He argued that ``semantic aspects of communication are irrelevant to the engineering problem. The significant aspect is that the actual message is one selected from a set of possible messages.''
We extend this idea to computation.

A communication channel that always emits the same symbol regardless of its input does not transmit any information;
likewise, any computation that maps every input to a fixed bit pattern is a `no-op' and is typically optimized away.
In contrast, transforming many possible inputs into many distinct outputs (such as arithmetic on encoded real numbers) is closer to our intuition about what represents meaningful computational work.

Shannon's \emph{mutual information}, $I(X;Y)$ as seen in \cref{eq:mutual_info}, quantifies computational work by measuring the amount of information shared between inputs and outputs.
It naturally captures bit widths, sparsity, noise, and potentially other hardware paradigms, as \cref{sec:benefits} suggests.
Mutual information's maximum over all input distributions is the \emph{compute-channel capacity} (\cref{eq:channel_capacity}), an application-agnostic, hardware-level performance limit.

We limit this section's scope to deterministic, digital computing for simplicity, with extensions left as future work.

\subsection{Encoding}\label{sec:encoding_proposal}
\begin{figure}
    \begin{tikzpicture}

\begin{axis}[
    height=\linewidth*0.8,
    ylabel={Probability},
    xlabel={Discrete Outcome},
    ymin=0, ymax=1.03,
    ytick={0, 0.1, 0.25, 0.5, 0.75, 1.0},
    xtick={1,2,3,4,5,6,7,8,9,10},
    enlarge x limits=0.05,
    ybar, 
]

\addplot+[
    color=red, fill=red, 
    fill opacity=0.5, draw opacity=1,
    bar shift=0pt
] coordinates {
    (1, 0) (2, 0) (3, 0) (4, 0) (5, 1)
    (6, 0) (7, 0) (8, 0) (9, 0) (10, 0)
};

\addplot+[
    color=blue, fill=blue, 
    fill opacity=0.5, draw opacity=1,
    bar shift=0pt
] coordinates {
    (1, 0.016054913299903592) (2, 0.043641779080613376) (3, 0.0923896470386652) (4, 0.15232477626512442) (5, 0.19558888431569338)
    (6, 0.19558888431569338) (7, 0.15232477626512442) (8, 0.0923896470386652) (9, 0.043641779080613376) (10, 0.016054913299903592)
}; 
 
\addplot+[
    color=teal, fill=teal, 
    fill opacity=0.5, draw opacity=1,
    bar shift=0pt
] coordinates {
    (1, 0.1) (2, 0.1) (3, 0.1) (4, 0.1) (5, 0.1)
    (6, 0.1) (7, 0.1) (8, 0.1) (9, 0.1) (10, 0.1)
};

\node[font=\bfseries, color=red, align=center] at (axis cs:2.5, 0.78) {Constant\\(H = 0.00 bits)};
\draw[->, red, thick] (axis cs:3, 0.85) -- (axis cs:4.5, 0.98);

\node[font=\bfseries, color=blue, align=center] at (axis cs:7.5, 0.42) {Normal\\(H $\approx$ 2.97 bits)};
\draw[->, blue, thick] (axis cs:7, 0.35) -- (axis cs:6.2, 0.22);

\node[font=\bfseries, color=teal, align=center] at (axis cs:2.5, 0.32) {Uniform\\(H $\approx$ 3.32 bits)};
\draw[->, teal, thick] (axis cs:2, 0.25) -- (axis cs:2, 0.12);

\end{axis}

\end{tikzpicture}
    \caption{
        A comparison of distributions and their entropies.
    }
    \label{fig:shannon_entropy}
\end{figure}
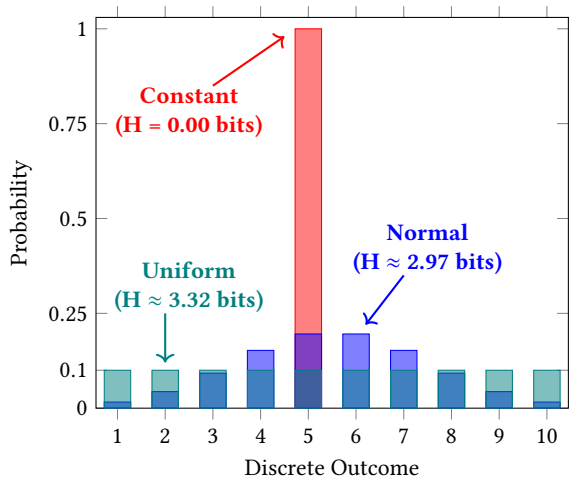

First, information is encoded into binary data through data types.
These define the representable input space, and Shannon entropy can be used to quantify the information captured by these data types.
In a discrete setting, Shannon entropy measures the expected information of a system's possible states by following \cref{eq:shannon_entropy} where $H(X)$ is the Shannon entropy of random variable $X$, $x$ is a potential value of $X$, and $p(x)$ is the probability of occurrence for $x$.
A few examples of distributions and their Shannon entropies are visualized in \cref{fig:shannon_entropy}.

As digital logic operates on binary values, we measure entropy using a base-2 logarithm, resulting in entropy units of `bits'.
This choice brings our computing performance metric \emph{back to bits}---just like communication.

\begin{equation}\label{eq:shannon_entropy}
    H(X) = - \sum_{x \in X} p(x) \text{log}_2p(x)
\end{equation}

Thus, the maximization of the input variables' entropies can be approximated by the bit width of the operands (achieved by a uniform distribution).
For operations with multiple input operands, a further approximation is to assume their independence.
The entropy of two independent variables is the sum of their individual entropies, so we can simply add the bit widths of all input operands.

For example, multiplying two uniformly distributed 8-bit integers\footnote{$p(x) = 1 / 2^8$ (a constant) and $|X|=2^8$.} gives a maximum input entropy of 16 bits.
This ideal approximates the U.S. Department of Commerce's current TPP metric~\cite{us_department_of_commerce_implementation_2023,us_department_of_commerce_framework_2025}.
However, our framework provides a first principles reasoning for this rather than ad hoc alterations.

Notably, these definitions capture the redundancy that some formats carry.
For instance, many formats use multiple bit patterns for not-a-number representations (NaNs), which reduces the effective entropy compared to the ideal (see \cref{sec:redundant_encodings}).
We can therefore define a data type's \emph{encoding efficiency} (or normalized entropy) as $\eta = \frac{H}{H_{\max}}$ where $H$ is the ideal Shannon entropy of all \emph{representable values} and $H_{\max}$ is the ideal entropy of all \emph{valid bit strings}.

\subsection{Operations}

\begin{figure}
    \definecolor{cleancolor}{RGB}{51,122,183}
\definecolor{noisycolor}{RGB}{217,83,79}
\definecolor{inputcolor}{RGB}{92,184,92}

\begin{tikzpicture}
\begin{axis}[
    width=\linewidth,
    height=\linewidth*0.65,
    xlabel={Value},
    ylabel={Probability},
    xmin=-280,
    xmax=280,
    grid=major,
    bar width=0.8,
    ybar,
    legend pos=north east,
    legend style={font=\normalsize},
]

\addplot[ draw=noisycolor] 
    table[x=bin_center, y=frequency, col sep=comma] {clean_c_xor_Int16_10000000_0.05.csv};
\addplot[ draw=cleancolor] 
    table[x=bin_center, y=frequency, col sep=comma] {clean_c_+_Int16_10000000_0.05.csv};

\legend{XOR, Add}

\end{axis}
\end{tikzpicture}
    \caption{
       The distributions of output variables for XOR (red) and addition (blue) of two uniformly-distributed, signed 8-bit integers stored as 16-bit integers based on 10 million samples.
    }
    \label{fig:hist_xor_add_int16}
\end{figure}
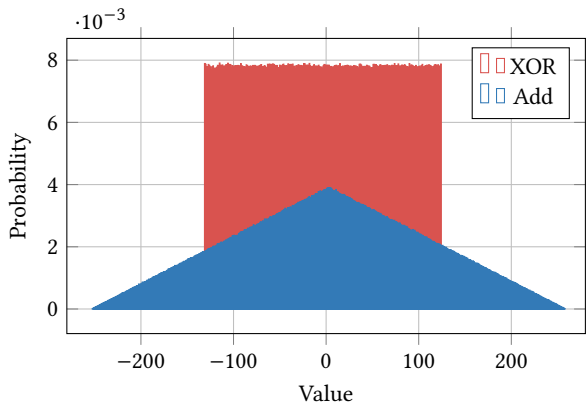

The encoded inputs are then transformed by a computation, such as addition, multiplication, or other logic.
This step may introduce information loss due to roundoff, overflow, underflow, NaN propagation, or how the mapping of input to output values behaves as shown in \cref{fig:hist_xor_add_int16} for integer XOR and addition.
In some settings, noise further reduces mutual information.
All of these effects appear as increases to the conditional entropy terms of mutual information, defined by \cref{eq:mutual_info} where $X$ and $Y$ represent input and output random variables, respectively.

\begin{equation}\label{eq:mutual_info}
    I(X;Y) = H(X) - H(X|Y) = H(Y) - H(Y|X) 
\end{equation}

\subsection{Decoding}

After transformation, the resulting outputs must be decoded, further constraining information flow.
Just as a communication channel's capacity is limited by its narrowest link, a computation's throughput is constrained by the entropy of its output representation.
Mutual information inherently captures this bottleneck: at best, it equals $\min(H(X),H(Y))$.

For example, adding two signed, 8-bit integers and storing the result in the same format results in a mutual information of less than 8 bits as expected. 
The value range of the sums exceeds the representable output range resulting in higher probabilities of the min and max output values.
Another example is a basic comparison, whose output is one of only two possible states, thus capping the mutual information to 1 bit.

\subsection{Compute-Channel Capacity}

Mutual information can assess performance for a specific input distribution or workload, but for hardware evaluation, there is a need for an application-agnostic upper limit.
For this purpose, we propose \emph{compute-channel capacity} as an information-theoretic computing performance unit.
As seen in \cref{eq:channel_capacity}, this unit quantifies the upper limit of information processed by a given hardware pipeline by maximizing over all input distributions.

\begin{equation}\label{eq:channel_capacity}
   C = \max_{p(x)}I(X;Y)
\end{equation}

Channel capacity (bits per channel use) converts to throughput (bit/s) by multiplying by channel use rate (e.g., clock frequency), analogous to deriving flop/s. For parallel systems, independent channel capacities/throughputs sum due to mutual information additivity.

This perspective offers a unifying framework to evaluate the performance of any information-processing system based on its fundamental ability to encode, manipulate, and retain information.

\section{Implications and Applications}\label{sec:benefits}

With the main components of our framework laid out, we now highlight a few realizations.
These include our proposal's ability to capture computing diversity, inform data type designs, and potentially unify analysis of communication and computation.

\subsection{Generalization across Data Types, Bit Widths, Sparsity, and Noise}

Mutual information makes it unnecessary to distinguish between flops, integer-ops, or other data types.
A unit based on quantities of information is invariant to how the binary strings are encoded or named.
They are all just bits.

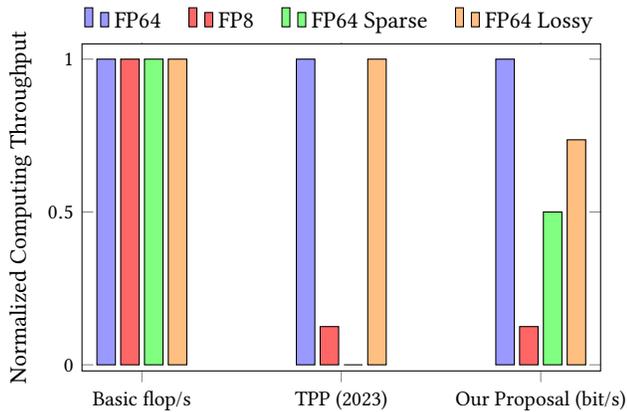
\begin{figure}
    \centering

\begin{tikzpicture}

\pgfplotscreateplotcyclelist{barstyles}{
  {fill=blue!40!white,     draw=black},  
  {fill=red!60!white,      draw=black},  
  {fill=green!50!white,    draw=black},  
  {fill=orange!50!white,   draw=black},  
  {fill=purple!40!white,   draw=black},  
  {fill=teal!50!white,     draw=black},  
  {fill=yellow!50!white,   draw=black},  
  {fill=gray!50!white,     draw=black},  
  {fill=brown!50!white,    draw=black}   
}

\begin{axis}[
    name=flopsplot,
    ybar,
    bar width=7pt,
    width=\linewidth,
    height=\linewidth*0.7,
    enlarge x limits=0.15,
    cycle list name=barstyles,
    ymax=1.05,
    ymin=-0.02,
    ylabel={Normalized Computing Throughput},
    symbolic x coords={flops_norm,tpp_2023_norm,ours_norm},
    xtick=data,
    xticklabels={Basic flop/s,TPP (2023),Our Proposal (bit/s)}, 
    legend style={
    at={(0.5,1.13)},        
    anchor=north,
    legend columns=4,
    draw=none,
  },
    tick label style={font=\small},
]

\addplot table[col sep=comma, x=Metric, y=FP64] {perf_compare_simple.csv};
\addplot table[col sep=comma, x=Metric, y=FP8] {perf_compare_simple.csv};
\addplot table[col sep=comma, x=Metric, y=FP64_sparse] {perf_compare_simple.csv};
\addplot table[col sep=comma, x=Metric, y=FP64_lossy] {perf_compare_simple.csv};

\legend{FP64\phantom{00},FP8\phantom{00},FP64 Sparse\phantom{00},FP64 Lossy}

\end{axis}

\end{tikzpicture}
    \caption{
       The reported computing throughput (normalized to FP64) for the following channels: FP64, FP8, FP64 with 50\% structured sparsity, and FP64 with a bit error ratio of 5\%.
    }
    \label{fig:simple_perf_bars_compares}
\end{figure}

More importantly, compute channel-capacity accounts for the bit widths of data types.
By contrast, simply counting flops means the number of FP64 operations would equal the number of 8-bit equivalents when swapping one type for the other, requiring additional qualification (``FP64 flops'' versus ``8-bit flops''), as illustrated in \cref{fig:simple_perf_bars_compares}.
Our approach is more in line with the U.S. government's export control approach with TPP.

In our metric, the maximum mutual information of any compute operation is bounded by the smaller of its total input entropy and its total output entropy.
Concretely, if an operation takes inputs of bit-widths ${bw_{\text{in},i}}$ and produces outputs of bit-widths ${bw_{\text{out},j}}$, then

\begin{equation}
    I_{\text{max}} \approx \min \left(\sum_i bw_{\text{in},i}  , \; \sum_j bw_{\text{out},j}\right).
\end{equation}

As a rough approximation for uniform input distributions, this reduces to TPP's input bit width accounting scheme.
When multiplied by the operation rate, channel capacity (in `bits per operation') thus provides a first-principles computing performance metric in line with a current government standard.
This approximation of mutual information mirrors practices in communication performance reporting, where bit error rate is often used as a proxy due to the challenges of calculating mutual information directly.

In communication systems, noise and interpretation of the channel as an identity operation justify such approximations.
Digital computing, however, is often lossless and considers many operations, yet similar simplifications remain useful.
Even when mutual information is difficult to compute explicitly, we believe practical approximations can still preserve its core theoretical motivations and enable practical performance assessment.

\subsubsection{Sparsity}

In traditional flop-based accounting, sparse operations are typically only counted for non-zero inputs.
For instance, the High Performance Conjugate Gradient (HPCG) benchmark reports only operations on non-zeros, which better reflects actual work performed~\cite{dongarra_new_2016}.
In contrast to these reporting practices, some hardware vendors support tensor operations that skip computations on structured sparse inputs yet still count these operations for performance evaluation~\cite{robert_crovella_dense_2024}.
The modern U.S. export control computing performance definition (TPP) sidesteps this discrepancy by explicitly considering only dense pipelines.

With our information-theoretic perspective, sparsity can be interpreted in terms of entropy.
A known zero entry (in structured sparsity or via a mask) carries zero entropy and requires no processing in an ideal case.
This makes intuitive sense.
If we are certain of the outcome, there is no information gain and thus no credit for performing the computation.
In contrast, an incidental zero value among the entire distribution carries information based on its log probability.

\subsubsection{Noise}
When noise affects a computation, the mutual information between inputs and outputs decreases (\cref{eq:mutual_info}), a loss our framework quantifies.
\Cref{fig:simple_perf_bars_compares} shows how random bit flips reduce the channel's structural fidelity, decreasing computational throughput.

\subsection{Redundant Encodings}\label{sec:redundant_encodings}

Recent efforts to eliminate redundant NaN encodings reflect an implicit priority: every bit pattern should carry unique information.
Unlike instruction-centric metrics such as Ryabko's and Fionov's~\cite{ryabko_estimating_2012}, our framework's basis in input/output entropies directly captures this effect.
As seen in \cref{sec:encoding_proposal}, redundancy reduces $\eta$ by allowing multiple encodings of the same value.

Our entropy-efficiency analysis helps explain why most modern formats drop redundant NaNs: they directly boost $\eta$.
The IEEE P3109 working group (Arithmetic Formats for Machine Learning) interim report mandates ``value sets shall include exactly one NaN,'' and the Open Compute Project's (OCP) low-precision specifications mostly adopt the same restriction~\cite{ieee_sa_p3109_working_group_interim_2025,rouhani_microscaling_2023}. 
The lone OCP exception, FP8\_E5M2 (5-bit exponent field and 2-bit mantissa field), retains redundancy but still achieves \(\eta=99.24\%\).
More strikingly, removing redundant NaNs in a 6-bit E2M3 increases $\eta$ from $86.12\%$ to $100\%$.

Meanwhile, a hypothetical FP8\_E4M3 with full IEEE-754 NaN redundancy would only reach \(\eta=97.40\%\), just below FP16's \(97.85\%\).
Given the focus on improving FP8 encoding efficiency and the similarity in $\eta$ between FP16 and FP8\_E4M3 with redundant NaNs, our entropy analysis suggests it may also be worth revisiting FP16's redundant NaN encodings.

\subsection{Communication and Computation}
Under our framework, communication is simply computation that performs an identity transformation---transferring data without altering its structure.
This view unifies computation and communication under a single performance paradigm.
Both are evaluated by the same metric: mutual information throughput (bit/s).
For example, our framework creates a dimensionally consistent roofline model~\cite{williams_roofline_2009}, transforming the x-axis from arithmetic intensity in flops per byte to a unitless bits of computation per bits of communication, and the y-axis from flops per second to bits per second.

Moreover, mutual information measures bits of \emph{information}, not raw, physical bits. 
Efficient compression or encoding increases information throughput at fixed bit-widths and clock rates, so two kernels with identical data widths can still differ in real performance. 
We call this factor \emph{information efficiency}, defined as the operational mutual information divided by compute-channel capacity.

Finally, by modeling memory hierarchies (e.g. L1, L2, RAM, interconnects, etc.) as stages in a composite channel, our approach offers a unified, information-theoretic performance analysis across scale capturing in-memory and in-network computing.

\begin{acks}
We thank Ben Hawkins for fruitful discussions, Ben Burns for his probability expertise, Luke Hawkins and Anna Freeman for their editing, and Ben Wilfong for his consistent camaraderie.

This research was supported in part through research infrastructure and services of the Rogues Gallery testbed~\cite{young_experimental_2019,powell_wrangling_2019} hosted by the Center for Research into Novel Computing Hierarchies (CRNCH) at Georgia Tech, through the National Science Foundation (NSF) Award Number \#2016701. 
Any opinions, findings and conclusions, or recommendations expressed in this material are those of the author(s), and do not necessarily reflect those of the NSF.
\end{acks}

\vspace{4em}

\bibliographystyle{ACM-Reference-Format}
\bibliography{zotero_my_library,references}




\end{document}